\documentclass{iopart}
\usepackage{iopams}
\usepackage{graphicx}
\usepackage{hyperref}
\usepackage{braket}

\begin{document}

\title{Winding number order in the Haldane model with interactions}

\author{E. Alba$^1$, J. K. Pachos$^2$, J. J. Garc{\'\i}a-Ripoll$^1$}
\address{$^1$ Instituto de F{\'\i}sica Fundamental IFF-CSIC, Calle Serrano 113b, Madrid 28006, Spain}
\address{$^2$ School of Physics and Astronomy, University of Leeds, Leeds LS2 9JT, UK}
\ead{jj.garcia.ripoll@csic.es}

\begin{abstract}
We study the Haldane model with nearest-neighbor interactions. This model is physically motivated by the associated ultracold atoms implementation. We show that the topological phase of the interacting model can be characterized by a physically observable winding number. The robustness of this number extends well beyond the topological insulator phase towards attractive and repulsive interactions that are comparable to the kinetic energy scale of the model. We identify and characterize the relevant phases of the model.
\end{abstract}


\section{Introduction}

The integer quantum Hall effect initially\ \cite{VonKlitzing1986} and the more extended set of topological insulators more recently\ \cite{Hasan2010}, represent a family of many-body systems with exotic transport properties. These properties originate from a robust and hidden topology that appears in the bands of the system\ \cite{Kohmoto1985}. More precisely, if such two-dimensional non-interacting models are embedded in a torus and diagonalized in momentum space, they give rise to a collection of bands, $\varepsilon_n(\mathbf{k})$, and eigenstates, $\psi_n(\mathbf{k})$, which are characterized by the Chern numbers
\begin{equation}
  \nu_n = \frac{i}{2\pi}\int \mathrm{d}^2\mathbf{k}
  \left[
  \partial_{k_x}\braket{\psi_n |\partial_{k_y}\psi_n}-
  \partial_{k_y}\braket{\psi_n |\partial_{k_x}\psi_n}
  \right] \in \mathbb{Z}.
\end{equation}
When the Fermi energy lays in the gap between two bands and the total Chern number of the occupied bands is nonzero, the resulting state is a topological insulator. Its key physical property is that it is non-conducting in the bulk, but it has robust conducting states that live on the boundary of the lattice. These edge states are well known for their robustness under perturbations: the nonlocal nature of the topological phase that supports them, allows edge states to survive under extreme conditions of distortion, impurities and other external perturbations. This resilience of topological edge states is highly favorable when considering their realization and detection in the laboratory.

An important open question in topological insulators is what happens to their topological properties in the presence of interactions. One special situation is when we force the energy bands of the insulator to be extremely narrow, having $\varepsilon_n(\mathbf{k})$ almost independent of $\mathbf{k}$, and introduce a strong repulsion or attraction. A paradigmatic case is the fractional quantum Hall effect (FQHE)\ \cite{Stormer1999}, where states with fractional filling present strong correlations and excitations with anyonic statistics that could be used for quantum computing\ \cite{Nayak2008}.

A complementary situation is when there are obvious dominant energy scales and the interaction term is comparable to the kinetic term that originates the topological phase. This regime, which has been addressed in earlier theoretical works\ \cite{Varney2010,Varney2011}, is relevant to what could be experimentally achieved in present quantum simulations of topological insulator with ultracold atoms. Such simulations have been recently realized in the laboratory for the Harper-Hofstadter Hamiltonian\ \cite{Aidelsburger2013,Miyake2013} and the Haldane model\ \cite{Jotzu2014}.

\begin{figure}[b]
  \includegraphics[width=0.8\linewidth]{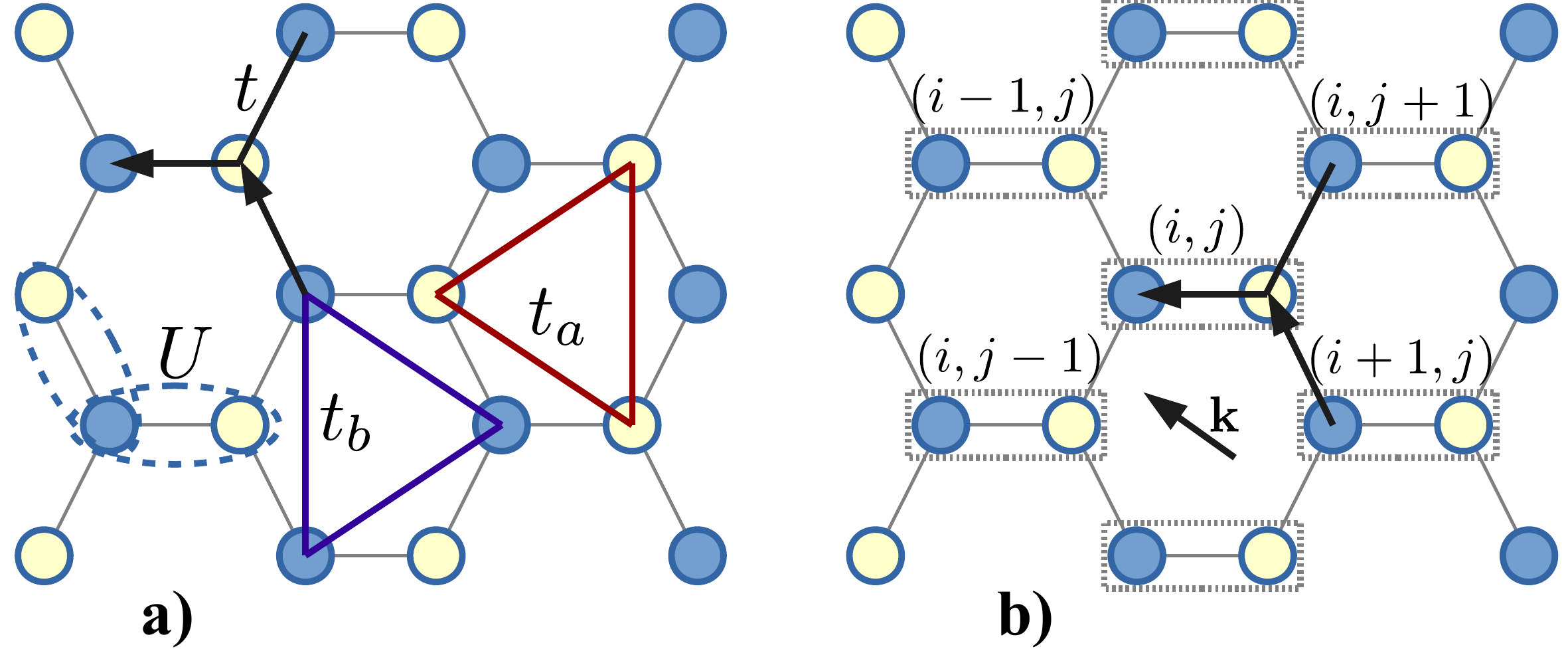}
  \caption{(a) Honeycomb lattice model depicting the nearest neighbor, $t$, next-nearest-neighbor hoppings, $t_{a,b}$, and the interaction $U$. (b) Lattice unit cell and indices, together with the orientation of the phase in the nearest neighbor hopping.}
  \label{fig:lattice}
\end{figure}

In this work we employ the Haldane model to study the role of interactions in the survival or destruction of the topological phases. This model\ \cite{haldane88} is a particular instance of a topological insulator or anomalous quantum Hall effect. It is realized on a honeycomb lattice with complex hopping amplitudes, as shown in \fref{fig:lattice}. We rely on a variant of the model introduced in Ref.\ \cite{alba13}, which is amenable to being implemented using ultracold atoms in optical lattices. It was also shown in Ref.\ \cite{goldman13} that it can introduce new regimes, such as a topological semimetal. This particular implementation naturally gives rise to nearest-neighbor interactions\ \cite{Alba2013}. This is exactly the setting we simulate here. 

Our study relies on an established tool for characterizing non-interacting topological phases, the so called \textit{winding number}. This is a mathematical object
\begin{equation}
  \nu = \frac{1}{4\pi}\int \mathrm{d}^2\mathbf{k}\,
  \mathbf{S}(\mathbf{k})\cdot[\nabla\mathbf{S}(\mathbf{k})
  \times \nabla\mathbf{S}(\mathbf{k})],
  \label{eq:winding}
\end{equation}
which can be constructed when the unit cell of the lattice is represented as a pseudospin and the eigenstates of the model are represented by unique Bloch vectors,
\begin{equation}
\mathbf{S}(\mathbf{k}) := \braket\bsigma = \mathrm{tr}(\bsigma \ket{\psi_0(\mathbf{k})}\bra{\psi_0(\mathbf{k})}).
\label{eq:spin-field}
\end{equation}
For the non-interacting case the winding number is identical to the Chern number\ \cite{DeLisle2014}. It has been shown that $\nu$ is an observable that can be directly measured in ultracold atoms experiments\ \cite{alba13}. This approach has been generalized to topological superconductors\ \cite{Pachos2013} and to composite problems (i.e. unit cell dimensionality $2^n$), where it still signals the existence of topological phase\ \cite{DeLisle2014}. In this respect, we regard the winding number as a global order parameter in its own right that can be measured experimentally.

The main result of this work is that the topological phase that originates in the non-interacting topological insulator regime extends towards both attractive and repulsive interactions, and is \textit{at all times detectable through the winding number}\ \eref{eq:winding}. For very repulsive interactions a charge density wave order (CDW) develops, as already indicated by exact diagonalizations in finite lattices with a fixed number of particles per site\ \cite{Varney2010,Varney2011}. For very attractive interactions the lattice fills up, as we work at zero chemical potential regime. This phenomenology can be observed with a very straightforward mean-field theory that works with the vector field\ \eref{eq:spin-field} as an order parameter. This theory shows both the survival of the winding number as well as the phase transitions into the topological order and into the CDW. These mean-field predictions are confirmed by Matrix-Product-States (MPS, also known as DMRG) ground state computations with up to 50 lattice sites, a size that doubles what is currently achieved through exact diagionalizations. Finally, the destruction of the topological insulator that takes place for large attractive interactions is modeled using an extension of the mean-field theory by Poletti {\em et al.}\ \cite{Poletti2011}, which shows the appearance of a superconducting phase in such regimes.

The structure of this paper is as follows. In section\ \ref{sec:model} we introduce the form of the Haldane model used here. Section\ \ref{sec:mean-field} introduces a mean-field theory that works with the spin field\ \eref{eq:spin-field} as order parameter and thus gives at all times a proper value of the winding number. Section\ \ref{sec:mps} introduces Matrix-Product-States, the variational ansatz that we use to get numerical estimates of the ground state properties in finite-size problems. Since both our mean field and our DMRG fail for very attractive interactions, section\ \ref{sec:bcs} introduces a BCS ansatz that we can use to describe that regime, generalizing the ideas in\ \cite{Poletti2011}. Section\ \ref{sec:results} introduces the main results, describing the phase transitions that take place in this model and how they are characterized through the winding number, the mean field and DMRG, or the BCS theory. Finally, section\ \ref{sec:discussion} discusses further implications of this work and possible outlooks.

\section{The model}
\label{sec:model}

We work with a variant of the Haldane model that introduces flux as a phase on the nearest neighbor hopping of a honeycomb lattice, as explained in Ref.\ \cite{alba13}. The model is extended to consider also nearest-neighbor interactions that take place between the different sublattices of the Haldane model. Overall, we can write the Hamiltonian in the form $H=H_0+H_U$, with the Haldane model describing the hopping of particles in the honeycomb lattice. Introducing pairs of indices, $\mathbf{s}=(i,j)$, running over the lattice, as shown in Fig.\ \ref{fig:lattice}, we can write
\begin{eqnarray}
  H_0 &=& -\sum_{\mathbf{v}\in\{\mathbf{v}_0,\mathbf{v}_1,\mathbf{v}_2\}} \sum_{\mathbf{s}} (t_{\mathbf{v}} b^\dagger_{\mathbf{s}+\mathbf{v}} a_{\mathbf{s}} + \mathrm{H.c.})
  - \sum_{\mathbf{s}} \epsilon (a^\dagger_{\mathbf{s}} a_{\mathbf{s}} - b^\dagger_{\mathbf{s}} b_{\mathbf{s}}) \\
  &&- \sum_{\mathbf{v}\in\{\mathbf{v}_1,\mathbf{v}_2,\mathbf{v}_3\}}\sum_{\mathbf{s}} (t_a a^\dagger_{\mathbf{s}+\mathbf{v}} a_{\mathbf{s}} + t_b b^\dagger_{\mathbf{s}+\mathbf{v}} b_{\mathbf{s}})\nonumber
\end{eqnarray}
and a nearest-neighbor interaction
\begin{equation}
  H_U = \sum_{\mathbf{v}\in\{\mathbf{v}_0,\mathbf{v}_1,\mathbf{v}_2\}}\sum_{\mathbf{s}}  Un_{a,\mathbf{s}}n_{b,\mathbf{s}+\mathbf{v}}
\end{equation}
Here, $\mathbf{v}_i\in\{(0,0),(1,0),(0,1),(1,-1)\}$ represent displacements between lattice cells; $t_{\mathbf{v}}=t\exp(i\phi_{\mathbf{v}})$, and the intralattice hoppings, $t_{a,b}$, as well as the nearest-neighbor interaction $U$ and a possible lattice imbalance, $\epsilon$, that will be made zero in this work.

The hopping part of the Hamiltonian, $t$ and $t_{a,b}$, implements a Dirac-type Hamiltonian with two distinct singularities in the Brillouin zone. The $t$ term gives rise to the kinetic terms of the effective model, while the $t_a$, $t_b$ terms implement a momentum dependent mass. The combination of both terms can be rewritten as a pseudospin model
\begin{equation}
  H_0 = -\int_{\mathrm{BZ}}\mathrm{d}^2\mathbf{k}\, \varepsilon(\mathbf{k})\mathbf{S}(\mathbf{k})\cdot[\Psi^\dagger(\mathbf{k})\bsigma\Psi(\mathbf{k})],
\end{equation}
where $\pm \varepsilon(\mathbf{k})$ are the energies of the two bands at the given quasimomentum $\mathbf{k}$. At half filling, that is one atom per unit cell, the ground state is obtained by placing one particle in the lowest band, which is a single particle state with an orientation of the pseudospin $\Psi(\mathbf{k})^\dagger = (a_{\mathbf{k}}^\dagger,b_{\mathbf{k}}^\dagger)$ dictated by $\mathbf{S}$.

As explained in Ref.\ \cite{alba13}, this model could be implemented using two separate optical lattices that store atoms in two different internal states, one for sublattice A and a different state for sublattice B. The nearest neighbor hopping would then be implemented by laser assisted tunneling between internal states, so that the hopping $t_{ij}$ would carry the phase difference between internal states at their respective positions, $t_{xy} \propto \exp[\mathbf{k}\cdot(\mathbf{x}+\mathbf{y})]$. In our setup we have chosen one fixed direction for the phase, $\mathbf{k}$, shown in \fref{fig:lattice}b, so that the complex hoppings can be written as
\begin{equation}
\label{eqn:complexflux}
  t_{\mathbf{v}_0} = e^{i\Phi} = t_{\mathbf{v}_1}^*.
\end{equation}
This approach has the characteristic that the overlap between neighboring sites, which makes it possible to have nearest-neighbor assisted tunneling, also allows for a strong nearest-neighbor interaction (c.f. Ref.\ \cite{Alba2013}).

\section{Mean field theory}
\label{sec:mean-field}

\subsection{General idea}

Our approach towards doing a mean-field study of the Haldane model builds on our knowledge of the non-interacting solution: in this case, the ground state wavefunction is exactly parameterized by the pseudospin orientation $\mathbf{S}(\mathbf{k})$ of the fermion that populates the $\mathbf{k}$ quasimomentum in the Brillouin zone. Consequently, we may write the ground state wavefunction at half filling as
\begin{equation}
  \ket{GS} = \otimes_{\mathbf{k}} \ket{\mathbf{S}(\mathbf{k})},
  \label{eq:mf}
\end{equation}
where the tensor product is taken over the Hilbert spaces of each quasimomentum mode.

In the interacting case we expect the nearest-neighbor repulsion to have a moderate effect on the fermion distribution, more or less preserving the Fermi sea and the topological order associated to the vector field $\mathbf{S}$. In other words, we will estimate variationally the ground state properties of the interacting model using the wavefunction\ \eref{eq:mf} and the orientations of the pseudospins as variational parameters.

This mean-field approach is quite unique in that we do not work with two variational parameters describing the unit cell in position space, but rather employs a quasi-continuous normalized vector field $\mathbf{S}(\mathbf{k})$ defined on a lattice of $N$ equispaced points in the Brillouin zone. Overall we have to work with $2^{N}$ degrees of freedom representing the orientation of the pseudospins in such sampling. It is important to remark that this method becomes exact in the $U=0$ limit. 

Moreover, the mean-field variational wavefunction\ \eref{eq:mf} continuously interpolates between the topologically ordered phase, and other phases that are to be found, such as the charge density wave (CDW) that we expect in the limit $U\to+\infty$, where atoms localize in either of the sublattices. This variational ansatz, however, does not capture other orders, such as a superconducting phase that should arise for large negative $U \sim -2 |t|$, which have to be analyzed using a different ansatz.

\subsection{Spin model}

We rewrite the Haldane model in momentum space, including the interaction. For that we introduce the Fourier transformed operators
\begin{equation}
  a_{\mathbf{k}} = \frac{1}{N^{1/2}}\sum_{\mathbf{s}}
  e^{-i \mathrm{k} \mathbf{s}} a_\mathbf{s},
\end{equation}
and similarly for the $b$ operators. Here $\mathbf{k}$ is a set of $N$ discrete momentum operators that span the first Brillouin zone (BZ), and $\mathbf{r}$ is the lattice position.

The interaction term is written in position space as
\begin{equation}
  H_U = U \sum_{\mathbf{v}_m}\sum_{\mathbf{s}}
  b^\dagger_{\mathbf{s}+\mathbf{v}_m}b_{\mathbf{s}+\mathbf{v}_m}
 a^\dagger_{\mathbf{s}}a_{\mathbf{s}},
\end{equation}
with three displacements $\mathbf{v}_{0,1,2}$ that connect one site to its neighboring cells. The Fourier transform of this Hamiltonian becomes
\begin{equation}
  H = \frac{U}{N}\sum_{\mathbf{k}_{1,2,3,4}}
  b^{\dagger}_{\mathbf{k}_1} b_{\mathbf{k}_2} a^{\dagger}_{\mathbf{k}_3} a_{\mathbf{k}_4}
  \delta(\mathbf{k}_1-\mathbf{k_2}+\mathbf{k}_3-\mathbf{k}_4)
  f(\mathbf{k}_1-\mathbf{k_2}).
\end{equation}
The central exponential, summed over $\mathbf{r}$ has lead to a $\delta(\cdot)$ that enforces the conservation of momentum. Finally, we have the weight function
\begin{equation}
  f(\mathbf{q}) = \sum_{\mathbf{v}_m} e^{i\mathbf{q}\cdot\mathbf{v}_m}.
\end{equation}

At this point we notice that $H_U$ can be decomposed into terms that connect two momenta and terms that connect four different momenta. Since we are going to use a variational wavefunction of the form\ \eref{eq:mf}, the latter terms do not contribute. Our Hamiltonian thus reads
\begin{equation}
  H = -\sum_{\mathbf{k}}\varepsilon(\mathbf{k})\mathbf{S}(\mathbf{k})\cdot\bsigma
  + \frac{U}{N}\sum_{p \neq q} \left[b^\dagger_{\mathbf{p}}b_{\mathbf{p}}
  a^\dagger_{\mathbf{q}}a_{\mathbf{q}} f(0)
  + b^\dagger_{\mathbf{p}}b_{\mathbf{q}}
  a^\dagger_{\mathbf{q}}a_{\mathbf{p}} f(\mathbf{p}-\mathbf{q})\right].
\end{equation}
Within our variational subspace of one particle per momentum orbital, we have the relations
\begin{equation}
  b^\dagger_{\mathrm{q}} b_{\mathrm{q}} = \frac{1}{2}(1+ \sigma_{\mathrm{q}}^z),\;\;
  a^\dagger_{\mathrm{q}} a_{\mathrm{q}} = \frac{1}{2}(1- \sigma_{\mathrm{q}}^z),\;\;
  b^\dagger_{\mathrm{q}} a_{\mathrm{q}} = \sigma^+_{\mathrm{q}}.
\end{equation}
This allows us to rewrite, up to global constants,
\begin{equation}
  H = -\sum_{\mathbf{k}}\varepsilon(\mathbf{k})\mathbf{S}(\mathbf{k})\cdot\bsigma
  - \frac{U}{4N}\left(\sum_{\mathbf{p}} \sigma^z_\mathbf{p}\right)^2
  + \frac{U}{N}\sum_{\mathbf{p}\neq\mathbf{q}}\sigma^+_{\mathbf{p}}\sigma^-_{\mathbf{q}}
  f(\mathbf{p}-\mathbf{q}).
\end{equation}
The second term induces charge density wave order, forcing all atoms to be on one sublattice or the other, while the third term scatters particles  to different momenta. Both terms counteract the effect of the magnetic field $\varepsilon(\mathbf{p})\mathbf{S}(\mathbf{p})$ that tries to enforce topological order.

\section{Matrix Product States ground state}
\label{sec:mps}

\subsection{General idea}

\begin{figure}[t]
  \includegraphics[width=0.6\linewidth]{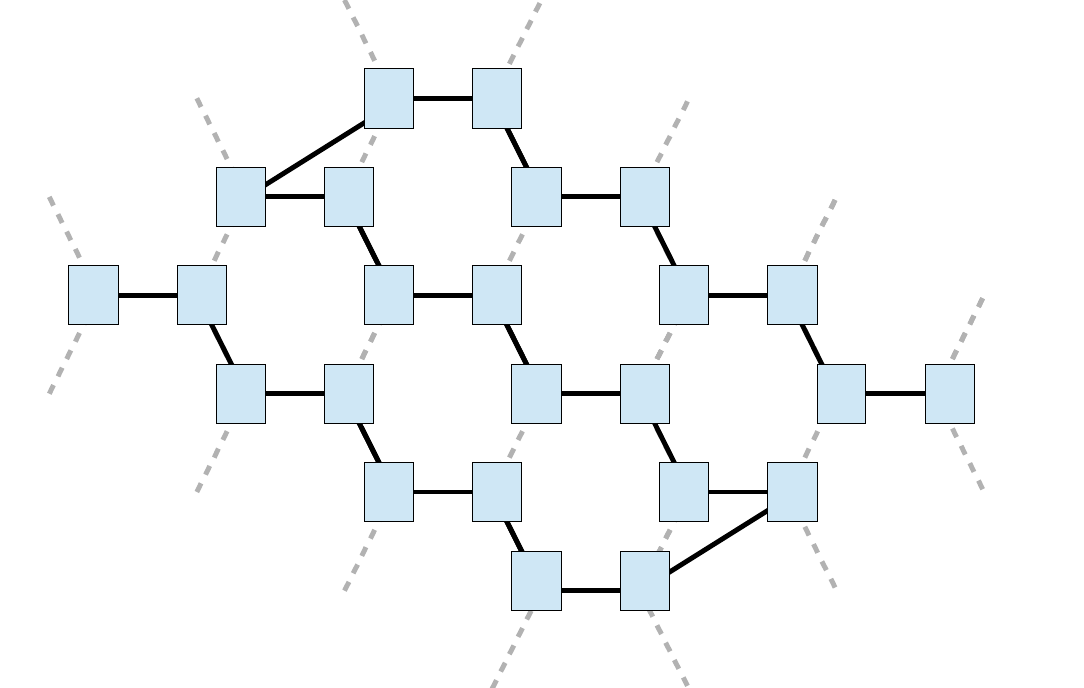}
  \caption{Matrix Product State structure running through the 2D honeycomb lattice. Dashed lines represent the underlying honeycomb lattice, while solid lines represent the MPS bond dimensions.}
  \label{fig:mps}
\end{figure}

One alternative to confirm the mean-field predictions would be to do some exact diagonalization to compute the ground state wavefunction, evaluating the observables that build up the winding number. However, this becomes unfeasible for two combined reasons.

The first one is that we need minimum size of the problem to detect the winding number. If we perform a simulation with $L_1\times L_2$ unit cells in position space, using periodic boundary conditions, that would amount to sampling $L_1\times L_2$ points in the Brillouin zone. A quick numerical calculation revealed that this sampling is insufficient to capture the winding number accurately if $L \le 3$. This means that the smallest simulation that could conceivable reproduce the non-interacting result would be a $4\times 4$ cells lattice with 32 sites.

The second reason is that given that baseline, performing simulations at half-filling becomes unfeasible using an ordinary computer and algorithms. Already storing the ground-state wave function for the $4\times 4$ lattice demands several gigabytes and we were not able to do ground state computations without resorting to state-of-the-art diagonalization software and a supercomputer with parallelized Lanczos. Note indeed that already the largest clusters that are found in the literature\ \cite{Varney2011} at half filling have sizes around 24 sites or $3\times 4$ cells, which are insufficient for the winding number computation.

A powerful alternative to exact diagionalizations are tensor-network state methods. In particular, we have used Matrix-Product-States to represent the ground state and a DMRG-type algorithm to compute the minimum energy state within that variational ansatz. An MPS is a variational form of a many-body wavefunction that is built by contracting tensors (matrices) in a 1D scheme. Using the ordering of lattice sites and tensors in \fref{fig:mps}, our ground state is written in the form
\begin{equation}
\ket{\Psi[A]} = \sum_{n_1\ldots n_N} \mathrm{tr}(A^{n_1}A^{n_2}\cdots A^{n_N})
\ket{n_1,n_2\ldots n_N},  
\label{eq:mps}
\end{equation}
where $n_m \in \{0,1\}$ is the occupation number of the $m$-th lattice site. Using a rather straightforward optimization procedure based on local updates of the tensors, it is possible to minimize the energy, computing the optimal tensors
\begin{equation}
  A = \mathrm{argmin}_A \frac{\braket{\Psi | H |\Psi}}{\braket{\Psi|\Psi}}.
\end{equation}

\subsection{Winding number from MPS}

After computing the ground state we still have to recover the winding number as measured in a time-of-flight experiment\ \cite{alba13}. In such experiments, the expansion of the atoms in the lattice makes them adopt a Fourier transform of their original wavefunctions\ \cite{pitaevskii},
\begin{eqnarray}
  a_i^\dagger \to \int \mathrm{d}^3\mathbf{p}\, \tilde{w}(\mathbf{p}) e^{i\mathbf{p} \mathbf{r}_{i}} \tilde\psi_a(\mathbf{p})^\dagger,\\
  b_i^\dagger \to \int \mathrm{d}^3\mathbf{p}\, \tilde{w}(\mathbf{p}) e^{i\mathbf{p} \mathbf{r}_{i}} \tilde\psi_b(\mathbf{p})^\dagger,
\end{eqnarray}
where the quasimomentum is mapped to the position of the expanding atoms, $\mathbf{p} = m \mathbf{r}/t$. In this formula $\mathbf{r}_{i}$ represents the center of the $i$-th lattice cell in the given sublattice, $s\in\{a,b\}$. The weight $\tilde{w}$ is the Fourier transform of the wavefunction of an atom trapped in one lattice site, and for deep enough lattices it can be taken constant over a large domain. Finally, $N$ is an overall normalization.

The previous expression implies that if we measure the spin texture of the expanding atoms
\begin{equation}
  \mathbf{v}(\mathbf{p}) = \braket{\Psi^\dagger(\mathbf{p}) \bsigma \Psi(\mathbf{p})},
\end{equation}
with $\tilde\Psi^\dagger = (\tilde\psi_a^\dagger, \tilde\psi_b^\dagger)$, this texture will be related to the spin texture of the original bands in the lattice
\begin{equation}
  \mathbf{v}(\mathbf{p}) \propto |\tilde{w}(\mathbf{p})|^2
  \sum_{\mathbf{m},\mathbf{n}} \braket{\Psi_{\mathbf{m}}^\dagger \bsigma \Psi_{\mathbf{n}}} e^{i\mathbf{p}(\mathbf{r}_{\mathbf{m}}-\mathbf{r}_{\mathbf{n}})}.
\end{equation}
In particular, when $\mathbf{p} = \mathbf{k}$ coincides with a quasimomentum in the Brillouin zone, $\mathbf{v}(\mathbf{k}) \propto \mathbf{S}(\mathbf{k})$, up to factors that drop out when we normalize $\mathbf{v}$ to recover $\mathbf{S}$.

Consistently with the previous reasoning, we have worked with the MPS wavefunction computing the vector field
\begin{equation}
  \mathbf{v}_{\mathbf{k}} = \frac{1}{M} \sum_{{\mathbf{m}},{\mathbf{n}}}\braket{\Psi_{\mathbf{m}}^\dagger \bsigma \Psi_{\mathbf{n}}} e^{i\mathbf{k}(\mathbf{r}_{\mathbf{m}}-\mathbf{r}_{\mathbf{n}})},
\end{equation}
where $\mathbf{k} \in \pi/M\times [-M/2,M/2)^{\otimes 2}$, is a finite sampling of the Brillouin zone. Using this sampling, we then compute the winding number using an accurate formula for the solid angle spanned by every three neighboring pseudospins, $\mathbf{S}=\mathbf{v}/|\mathbf{v}|$, on the momentum space lattice\ \cite{oosterom83},
\begin{equation}
  \nu = \sum_{\langle k,p,q\rangle} \Omega(\mathbf{S}_{\mathbf{k}},\mathbf{S}_{\mathbf{p}},\mathbf{S}_{\mathbf{q}}),
\end{equation}
with the solid angle approximation
\begin{equation}
  \tan\left[\frac{\Omega(\mathbf{a},\mathbf{b},\mathbf{c})}{2}\right]
  = \frac{\mathbf{a}\cdot(\mathbf{b}\times\mathbf{c})}
{|\mathbf{a}||\mathbf{b}||\mathbf{c}|+
(\mathbf{a}\cdot\mathbf{b})|\mathbf{c}|+
(\mathbf{a}\cdot\mathbf{c})|\mathbf{b}|+
(\mathbf{b}\cdot\mathbf{c})|\mathbf{a}|},
\end{equation}
that is valid for small $\Omega$.

\subsection{MPS technical remarks}

There are several important technical remarks regarding the MPS simulations. The first one regards the use of conserved quantities to restrict the simulation, imposing for instance, a total number of particles in the lattice. We have chosen not to do this, looking for the ground state of the full Hamiltonian, which is equivalent to minimizing the free energy at zero chemical potential. This means that in these simulations an increase or decrease of the filling is possible, though, as we will see below, it is irrelevant for the topological order as detected by the winding number --a signature of the robustness of this quantity, as shown already in\ \cite{alba13}.

The second remark regards the size and type of lattice. MPS simulations were done for lattices with $4\times 4$, $4\times 5$ and $5\times 5$ cells (32, 40 and 50 sites), the latter showing the clearest signal. We explicitly use open boundary for the MPS wavefunction because of several reasons. First of all, open boundaries are the ones that best reflect a potential experiment with ultracold atoms. Second, we wish to show that the winding number measurement is a robust and powerful observable that does not restrict to abstract geometries, such as tori and spheres. Third, since we already have periodic boundaries in the mean field theory, agreement between this theory and the OBC MPS represents a strong signature that our approach is correct.

Finally, we have to comment on the size of the MPS, the so called \textit{bond dimension}. This dimension is the size of the matrices in\ \eref{eq:mps} and it relates to the maximum amount of entanglement that is available in a bipartition of the state. The simulations that we show here are done with bond dimension $\chi=100$. This restriction is based on the need to scan in detail the whole parameter space and the use of long-range interactions induced by the 2D-to-1D mapping. However, in this particular study, for the observables that we computed, including density correlations and the full winding number, we have verified that convergence starts already at very low bond dimensions, such as 30 for the $5\times 5$ lattice. We believe that this is possible because PEPS can already represent topological insulator phases and topologically ordered states faithfully\ \cite{Yang2015} with moderate bond dimensions. Since PEPS may be rewritten as MPS with a linear increase in the bond dimension ($\chi \to \chi\times L$, where $L$ is the lattice diameter), our MPS representation, which is easier to operate numerically, does not need to be too complex to reproduce many of the relevant physical properties.

\section{BCS Ansatz}
\label{sec:bcs}

We can further analyze our system, in the case of attractive ($U<0$)
interactions, by means of a  using a BCS ansatz. In order to do so,
and to facilitate of comparison, we will follow the procedure used in
Ref. \cite{Poletti2011}, which deals with a similar tight-binding Hamiltonian.
Up to trivial factors, the Hamiltonian presented there corresponds to
the $\Phi=0$ case in Eq.~(\ref{eqn:complexflux}). The BCS ansatz is
based on a Wick expansion of the interaction term (overwrite term here) in momentum space which gives
\begin{equation} 
\hat{H}_{int} =U\sum_k \left(
  \Delta^\star_{\mathbf{k}}\hat{b}_{-\mathbf{k}}\hat{a}_{k}+\Delta_{\mathbf{k}}\hat{a}^\dagger_{\mathbf{k}}\hat{b}^\dagger_{-{\mathbf{k}}}-\Delta_{\mathbf{k}}\langle
  \hat{a}^\dagger_{\mathbf{k}}\hat{b}^\dagger_{-{\mathbf{k}}}\rangle \right),
\end{equation}
where
$\Delta_{\mathbf{k}}=-1/N^2\sum_{\mathbf{k}'}f(\mathbf{k}-\mathbf{k}') \langle
\hat{b}_{-\mathbf{k}'}\hat{a}_{\mathbf{k}'}\rangle$
is computed through a self-consistent calculation. The order
parameters which characterize the phase are nearest-neighbor two
particle pairing correlators
\begin{equation}
  \bdelta = \left(
    \braket{\hat{a}_{\mathbf{s}} \hat{b}_{\mathbf{s}+\mathbf{v}_0}}, \braket{\hat{a}_{\mathbf{s}} \hat{b}_{\mathbf{s}+\mathbf{v}_1}},
    \braket{\hat{a}_{\mathbf{s}} \hat{b}_{\mathbf{s}+\mathbf{v}_2}} \right),
    \label{eq:bcs}
\end{equation}
where $v_m$ are again the directions spanned by the three nearest
neighbors. Following Ref.\ \cite{Poletti2011}, we rewrite the
order parameter as $\bdelta/|\bdelta|=\sum_mw_m{\mathbf{u}}_m$, where the
$\mathbf{u}_m$ are a basis for the $\mathcal{S}_3$ cyclical
permutation group: $\mathbf{u}_1=(1,1,1)/\sqrt{3}$,
$\mathbf{u}_2=(2,-1,-1)/\sqrt{6}$ and
$\mathbf{u}_3=(0,1,-1)/\sqrt{3}$. A non-zero value of any $w_m$
coefficient (that is, of $|\bdelta|$) signals a superfluid phase,
while the particular coefficient renders information about the
symmetry of that phase. As we will see below, the results that we obtain in the attractive regime are consistent with those of Poletti et al\ \cite{Poletti2011}.

\section{Results}
\label{sec:results}

\begin{figure}[t]
  \includegraphics[width=0.9\linewidth]{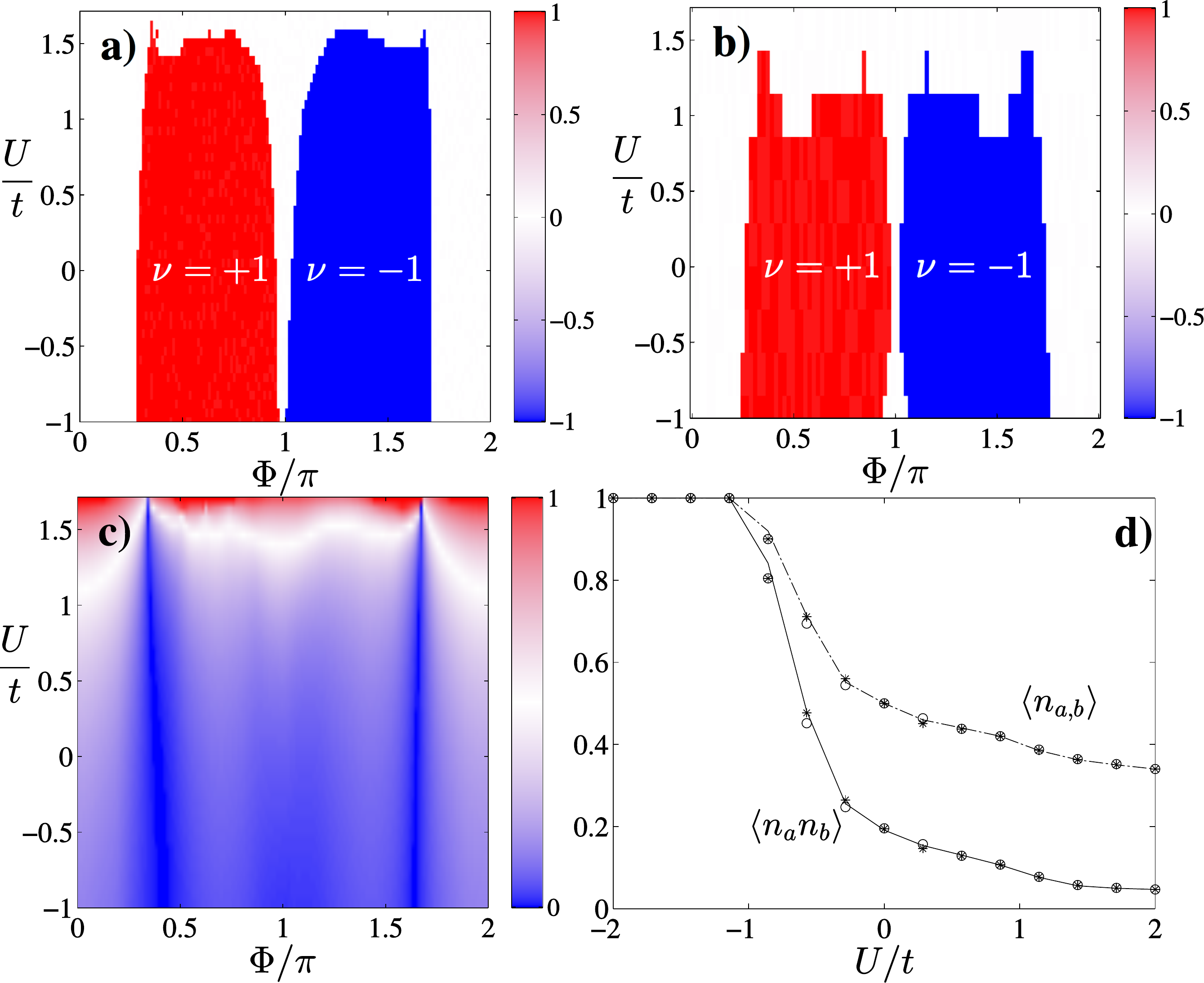}
  \caption{(a-b) Winding number as a function of the imparted phase $\Phi$ and the nearest neighbor interaction, $U$, using $t=1, t_a=-t_b=0.1$. We plot the outcome of the mean-field calculation (a) and of a MPS simulation (b) with $5\times5$ unit cells (50 sites), either in momentum (a) or in position space (b). (c) Mean field average value of $|\langle \sigma^z\rangle|$. (d) MPS expected value of the phase separation, $\langle n_{a,\mathbf{x}} n_{b,\mathbf{x}}\langle$ (solid), and of the average number of particles per site, $\langle n_{a,b}\rangle$ (dash-dot), as a function of the interaction, $U/t$, for various fluxes $\Phi/\pi=0$ (line), $0.5$ (star) and $1.5$ (circle).}
  \label{fig:mps}
\end{figure}

\Fref{fig:mps} summarizes the main results from the previous numerical methods. In figures \ref{fig:mps}a-b we plot the winding number of the mean-field or MPS simulation of the ground state. Both figures show a strong qualitative agreement, exhibiting a phase transition from a trivial phase around zero flux, $\Phi=0$, into a topological phase at larger fluxes where the Haldane phase is obtained for a wide range of interactions.

Along the vertical axis we find that the topological phase described by the winding number disappears for repulsive interactions around $U\sim t$. As shown in \fref{fig:mps}c, this transition is due to a spontaneous symmetry breaking of the $\sigma^z$ expectation values, signaling the transition into a charge density wave where particles are polarized into one sublattice or the other.

\begin{figure}[t]
  \includegraphics[width=0.8\linewidth]{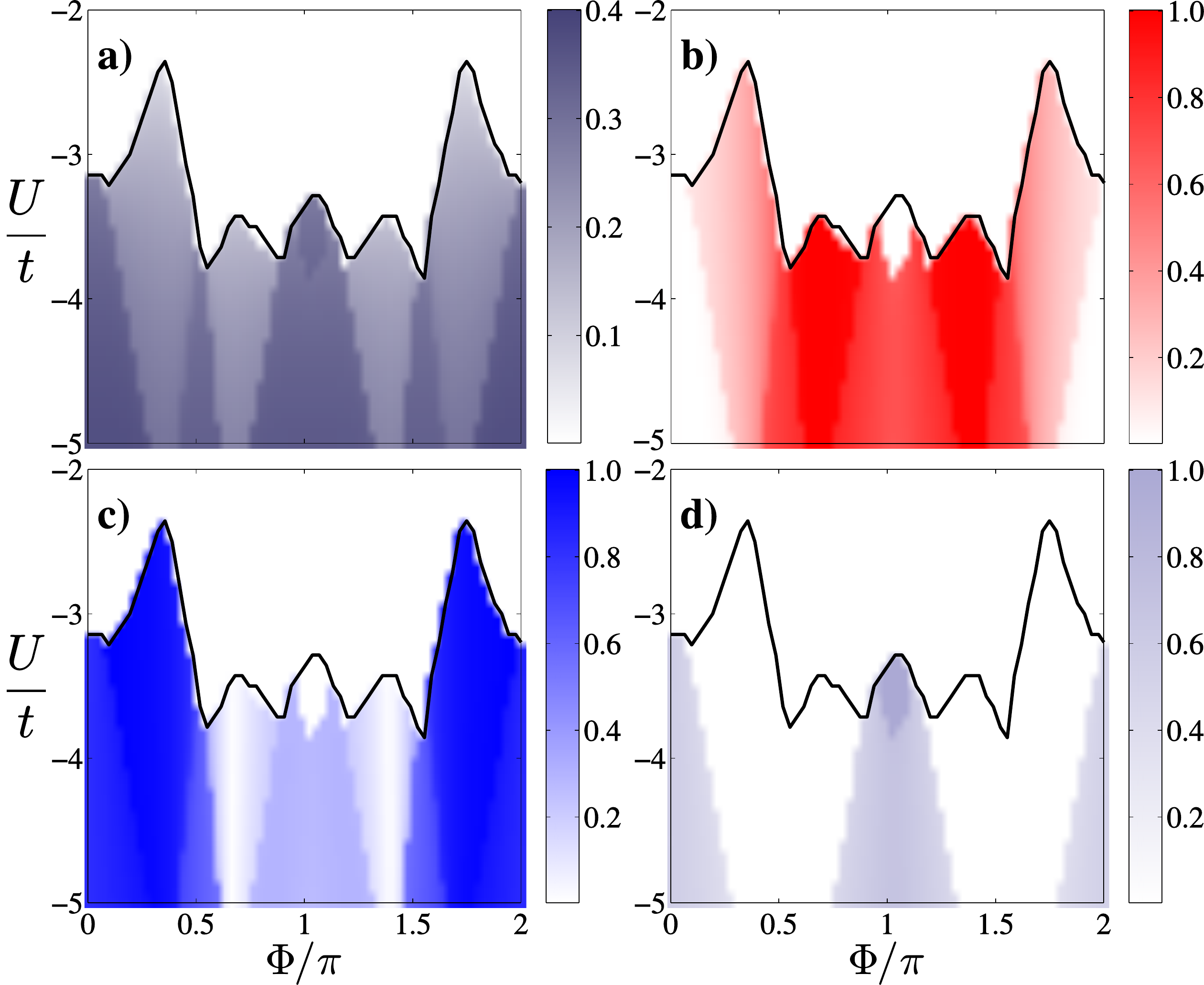}
  \caption{Results from the BCS ansatz. In figures (b-d) we plot the absolute value of the order parameters, $w_k$ for $k=1,2,3$, from the BCS ansatz\ \eref{eq:bcs}, while figure (a) shows the the norm $|\bdelta|$. The solid line delimits where the superconducting order parameter appears. Note that this happens well below the line $U=-1$ where our winding number simulations fail.}
  \label{fig:bcs}
\end{figure}

To confirm the phase separation or CDW order, we have studied a similar order parameter in the MPS simulation. In particular, we have computed the expectation value $\braket{n_{a\mathbf{s}} n_{b\mathbf{s}}}$, which measures the coexistence of atoms in neighboring sites in the honeycomb lattice. For large $U$ this value  decreases continuously down to zero, confirming the separation of species in the lattice. Interestingly, from the point of view of the CDW order, this looks like a cross-over, but from the point of view of the winding number this looks like sharp phase transition into a disordered regime. Whether this is an artifact of poor resolution on the numerical side (for instance because the norm of $\mathbf{v}$ becomes so small that our computation of the winding number is inaccurate), is something that our MPS simulations cannot resolve.

We have also studied what happens for negative or attractive interactions. When $U<0$, the ground state configuration at zero chemical potential has a filling fraction larger than $1/2$, that is more than one particle per unit cell. Despite this enlarged filling, the ordered phase still persists until $U=-1$, as evidenced by the winding number [\Fref{fig:mps}a-b]. At this critical interaction the lattice becomes perfectly filled [\Fref{fig:mps}d], forming a Mott insulator with two particles per site. This trivial configuration cannot be reproduced with the mean-field, because that wavefunction assumed half-filling. In order to study this region of strong attractions we have to use the BCS ansatz developed above\ (\ref{sec:bcs}). Figures \ref{fig:bcs}a-d show the outcome of that ansatz, revealing that the superconductor phase that was predicted for the honeycomb lattice\ \cite{Poletti2011} exists also in the full Haldane model. Most important, the symmetry of that phase is strongly dependent on the breaking of time-reversal symmetry in the non-interacting limit, i.e. of the topological
phase of the $U=0$ Hamiltonian.

\section{Discussion}
\label{sec:discussion}

Summing up, our study reveals that the topological insulator phase survives for a wide range of interaction and that this phase is faithfully detected by the winding number operator. Given that earlier exact diagionalizations computed the Chern number in selected regimes of parameters\ \cite{Varney2010,Varney2011}, this leads us to conclude that the winding number can also reproduce the Chern number in interacting systems with moderate correlations.

The ideas put forward in this work have very straightforward extensions to other models, including other topological insulators and composite systems\ \cite{DeLisle2014}, and other types of interactions, such as on-site Hubbard terms or long-range interactions. In all these systems it would be interesting to see whether the winding number may act as a precursor of other strongly correlated phases.

One may also wonder about the applicability of our results to the recent beautiful experiment demonstrating the Haldane model in an optical lattice\ \cite{Jotzu2014}. This experiment uses laser assisted tunneling to implement $t_{a,b}$, relying on the original lattice to supply $t$. Compared with our original proposal\ \cite{alba13}, it has the problem that the small overlap between neighboring sites would lead to a small value of $U$. In this case it would be more advantageous to implement other types of interactions, such as relying on two-level atoms and implementing on-site repulsion or attraction. Such models fall out of the scope of this work, but could be studied using a generalization of our MPS and mean-field methods above, combined with the study of partial winding numbers\ \cite{DeLisle2014} and how they relate to the global topological order.

Finally, we would like to emphasize the mean-field ansatz\ \ref{eq:mf}, which could be of interest to other contexts and models. Such momentum-representation mean-field theory differs from other mean-field theories that have been developed in position space\ \cite{Dauphin2012}, and connect to long-range interaction classical spin models that have been long analyzed in the literature. The connection between these models, the symmetries of the underlying topological insulator and how these are broken by the interaction terms could be a profitable avenue to understand the nature of the phase transitions that have been found in this study.

\ack

The authors acknowledge useful discussions with Zlatko Papic. This work has been supported by Spanish MINECO Project FIS2012-33022 and CAM Research Network QUITEMAD+.

\section*{References}


\begin{thebibliography}{10}

\bibitem{VonKlitzing1986}
Klaus von Klitzing.
\newblock {The quantized Hall effect}.
\newblock {\em Reviews of Modern Physics}, 58(3):519--531, July 1986.

\bibitem{Hasan2010}
M.~Z. Hasan and C.~L. Kane.
\newblock {Colloquium: Topological insulators}.
\newblock {\em Reviews of Modern Physics}, 82(4):3045--3067, November 2010.

\bibitem{Kohmoto1985}
Mahito Kohmoto.
\newblock {Topological invariant and the quantization of the Hall conductance},
  April 1985.

\bibitem{Stormer1999}
Horst Stormer, Daniel Tsui, and Arthur Gossard.
\newblock {The fractional quantum Hall effect}, March 1999.

\bibitem{Nayak2008}
Chetan Nayak, Steven~H. Simon, Ady Stern, Michael Freedman, and Sankar {Das
  Sarma}.
\newblock {Non-Abelian anyons and topological quantum computation}.
\newblock {\em Reviews of Modern Physics}, 80(3):1083--1159, September 2008.

\bibitem{Varney2010}
Christopher~N. Varney, Kai Sun, Marcos Rigol, and Victor Galitski.
\newblock Interaction effects and quantum phase transitions in topological
  insulators.
\newblock {\em Phys. Rev. B}, 82:115125, Sep 2010.

\bibitem{Varney2011}
Christopher~N. Varney, Kai Sun, Marcos Rigol, and Victor Galitski.
\newblock {Topological phase transitions for interacting finite systems}.
\newblock {\em Physical Review B}, 84(24):241105, December 2011.

\bibitem{Aidelsburger2013}
M.~Aidelsburger, M.~Atala, M.~Lohse, J.~T. Barreiro, B.~Paredes, and I.~Bloch.
\newblock {Realization of the Hofstadter Hamiltonian with Ultracold Atoms in
  Optical Lattices}.
\newblock {\em Physical Review Letters}, 111(18):185301, October 2013.

\bibitem{Miyake2013}
Hirokazu Miyake, Georgios~A. Siviloglou, Colin~J. Kennedy, William~Cody Burton,
  and Wolfgang Ketterle.
\newblock {Realizing the Harper Hamiltonian with Laser-Assisted Tunneling in
  Optical Lattices}.
\newblock {\em Physical Review Letters}, 111(18):185302, October 2013.

\bibitem{Jotzu2014}
Gregor Jotzu, Michael Messer, R\'{e}mi Desbuquois, Martin Lebrat, Thomas
  Uehlinger, Daniel Greif, and Tilman Esslinger.
\newblock {Experimental realisation of the topological Haldane model}.
\newblock {\em Nature}, 515(7526):14, November 2014.

\bibitem{haldane88}
F.~D.~M. Haldane.
\newblock Model for a quantum hall effect without landau levels:
  Condensed-matter realization of the "parity anomaly".
\newblock {\em Phys. Rev. Lett.}, 61:2015--2018, Oct 1988.

\bibitem{alba13}
E.~Alba, X.~Fernandez-Gonzalvo, J.~Mur-Petit, J.~K. Pachos, and J.~J.
  Garcia-Ripoll.
\newblock Seeing topological order in time-of-flight measurements.
\newblock {\em Phys. Rev. Lett.}, 107:235301, Nov 2011.

\bibitem{goldman13}
N~Goldman, E~Anisimovas, F~Gerbier, P~Öhberg, I~B Spielman, and G~Juzeliūnas.
\newblock Measuring topology in a laser-coupled honeycomb lattice: from chern
  insulators to topological semi-metals.
\newblock {\em New Journal of Physics}, 15(1):013025, 2013.

\bibitem{Alba2013}
E.~Alba, X.~Fernandez-Gonzalvo, J.~Mur-Petit, J.J. Garcia-Ripoll, and J.K.
  Pachos.
\newblock {Simulating Dirac fermions with Abelian and non-Abelian gauge fields
  in optical lattices}.
\newblock {\em Annals of Physics}, 328(null):64--82, January 2013.

\bibitem{DeLisle2014}
James de~Lisle, Suvabrata De, Emilio Alba, Alex Bullivant, Juan~J
  Garcia-Ripoll, Ville Lahtinen, and Jiannis~K Pachos.
\newblock {Detection of Chern numbers and entanglement in topological
  two-species systems through subsystem winding numbers}.
\newblock {\em New Journal of Physics}, 16(8):14, August 2014.

\bibitem{Pachos2013}
Jiannis~K. Pachos, Emilio Alba, Ville Lahtinen, and Juan~J. Garcia-Ripoll.
\newblock {Seeing Majorana fermions in time-of-flight images of staggered
  spinless fermions coupled by s-wave pairing}.
\newblock {\em Physical Review A - Atomic, Molecular, and Optical Physics},
  88(1):013622, July 2013.

\bibitem{Poletti2011}
D.~{Poletti}, C.~{Miniatura}, and B.~{Gr{\'e}maud}.
\newblock {Topological quantum phase transitions of attractive spinless
  fermions in a honeycomb lattice}.
\newblock {\em EPL (Europhysics Letters)}, 93:37008, February 2011.

\bibitem{pitaevskii}
L.~Pitaevskii and S.~Stringari.
\newblock {\em Bose-Einstein Condensation}.
\newblock Oxford Univ. Press, Oxford, UK, 2008.

\bibitem{oosterom83}
A.~van Oosterom and J.~Strackee.
\newblock The solid angle of a plane triangle.
\newblock {\em Biomedical Engineering, IEEE Transactions on},
  BME-30(2):125--126, Feb 1983.

\bibitem{Yang2015}
Shuo Yang, Thorsten~B. Wahl, Hong-Hao Tu, Norbert Schuch, and J.~Ignacio Cirac.
\newblock {Chiral Projected Entangled-Pair State with Topological Order}.
\newblock {\em Physical Review Letters}, 114(10):106803, March 2015.

\bibitem{Dauphin2012}
A.~Dauphin, M.~M\"{u}ller, and M.~A. Martin-Delgado.
\newblock {Rydberg-atom quantum simulation and Chern-number characterization of
  a topological Mott insulator}.
\newblock {\em Physical Review A}, 86(5):053618, November 2012.

\end{thebibliography}

\end{document}